\documentclass[a4paper,11pt]{article}
\usepackage{pos}
\usepackage{bbold}
\usepackage{caption}
\usepackage{subcaption}
\usepackage{hyperref}

\newcommand{\SU}{\mathrm{SU}}

\newcommand{\one}{\mathbb{1}}

\newcommand{\Tr}{\mathrm{Tr}}

\title{Lattice Gauge Symmetry in Neural Networks}

\author{Matteo Favoni}
\author{Andreas Ipp}
\author*{David I.~M\"uller}
\author{Daniel Schuh}

\affiliation{
Institute for Theoretical Physics, TU Wien \\
Wiedner Hauptstrasse 8-10/136, Tower B, 1040 Wien, Austria}

\emailAdd{favoni@hep.itp.tuwien.ac.at}
\emailAdd{ipp@hep.itp.tuwien.ac.at}
\emailAdd{dmueller@hep.itp.tuwien.ac.at}
\emailAdd{schuh@hep.itp.tuwien.ac.at}

\abstract{
We review a novel neural network architecture called lattice gauge equivariant convolutional neural networks (L-CNNs), which can be applied to generic machine learning problems in lattice gauge theory while exactly preserving gauge symmetry. We discuss the concept of gauge equivariance which we use to explicitly construct a gauge equivariant convolutional layer and a bilinear layer. The performance of L-CNNs and non-equivariant CNNs is compared using seemingly simple non-linear regression tasks, where L-CNNs demonstrate generalizability and achieve a high degree of accuracy  in their predictions compared to their non-equivariant counterparts. 
}

\FullConference{%
 The 38th International Symposium on Lattice Field Theory, LATTICE2021
  26th-30th July, 2021
  Zoom/Gather@Massachusetts Institute of Technology
}

\begin{document}
\maketitle

\section{Introduction}

The concept of symmetry or equivariance under symmetry transformations is at the theoretical foundation of modern physics, and it is hard to overstate its importance. Noether's first theorem establishes a clear relationship between invariance of Lagrangians under continuous global symmetries and the existence of conserved quantities and conserved currents \cite{Noether:1918}. Global symmetries, as the name implies, are transformations that are applied the same way at every point in space time. In mechanical systems and field theories, energy and momentum conservation laws follow from invariance under space-time translations, whereas rotational invariance implies the conservation of angular momentum. More generally, global symmetry under the Poincar\'{e} group, which includes translations, rotations and boosts, is the foundation of special relativity. In addition, symmetries not associated with isometries of space-time are of particular importance to quantum field theory. For example, global U(1) invariance in field theories of fermions and complex scalars implies the existence of locally conserved particle currents and globally conserved particle numbers. On the other hand, local symmetry, associated with continuous differentiable transformations that are functions on space time, is the foundation of gauge theories. For example, quantum chromodynamics (QCD) is a gauge theory with symmetry group $\SU(3)$. All known fundamental forces are formulated as gauge theories and therefore associated with particular local symmetries or gauge groups. 

Machine learning methods can make use of the concept of symmetry in a similar manner. Exploiting the geometry and symmetries of a particular machine learning problem to develop appropriate neural network architectures is the main idea behind geometric deep learning (see recent reviews \cite{Bronstein:2021aaa} and \cite{Gerken:2021sla} for a mathematical introduction). Convolutional neural networks (CNNs) can be understood as special cases of generic neural networks that respect translational symmetry. In the past decades, these types of neural networks have proven to be exceptionally useful in image recognition and computer vision in general. More specifically, CNNs are neural networks consisting of convolutional layers acting on image data or feature maps $f: \mathbb{Z}^2 \rightarrow \mathbb{R}^n$, where $\mathbb{Z}^2$ is the base space on which the image is defined and $n \in \mathbb{N}$ denotes the number of channels (e.g.~$n=3$ for an RGB image). Convolutional layers are equivariant in the sense that a spatial translation of the input feature map induces an appropriate translation of the output feature map. Equivariance of CNNs naturally lends itself to applications in lattice field theory, which typically exhibit translational symmetry \cite{Bulusu:2021rqz}. Going beyond translations, the framework of group equivariant convolutional neural networks (G-CNNs) \cite{Cohen:2016aaa} generalizes the equivariance property to a general group $G$ with the traditional CNN being a special case when $G$ is identified with the translation group. Apart from their theoretical appeal, G-CNNs with rotational symmetry have shown to be more robust on certain image recognition tasks, where traditional CNNs can fail to make correct predictions when provided with previously unseen rotated images. In the parlance of physics, G-CNNs are neural networks exhibiting global symmetry. A generalization of equivariant neural networks to local symmetries has been proposed in Ref.~\cite{Cohen:2019aaa} called gauge equivariant CNNs. In this architecture, the base space on which the data is defined is generalized to a curved manifold, as opposed to a flat base space. In order to retain equivariance in convolutional layers, feature maps must be appropriately parallel transported in order to obtain the correct transformation behavior under coordinate or frame changes. Although the term ``gauge equivariant'' would, to physicists, imply gauge symmetry in the sense of gauge theories such as electrodynamics or QCD, parallel transport of feature maps in these types of networks is purely due to the curved geometry of the base space and not due to the presence of a gauge field. Recently, several groups have tackled the problem of using machine learning methods in the context of lattice QCD or pure lattice gauge theory, while retaining gauge symmetry in the sense of gauge theories. For example, gauge equivariant normalizing flows \cite{Kanwar:2020xzo, Boyda:2020hsi, Albergo:2021vyo} have been successfully used to train generative models which can produce statistically independent lattice configurations for Monte Carlo simulations. Gauge covariant neural networks that can perform smearing and Wilson flow have also been studied \cite{Tomiya:2021ywc}. However, a general framework akin to G-CNNs or gauge equivariant CNNs for lattice gauge theory has been lacking so far. 

Lattice gauge theory is an exactly gauge covariant formulation of non-Abelian Yang-Mills theory discretized on a lattice originally proposed by Wilson \cite{Wilson:1974sk}.  The degrees of freedom are gauge link variables $U_{x,\mu} \in \mathrm{SU}(N_c)$ defined on the edges of a hypercubic lattice $\Lambda$ with lattice spacing $a$. A gauge link $U_{x,\mu}$ is defined along the edge $(x,x+\mu)$ starting at the lattice site $x \in \Lambda$ and ending at $x+\mu$, which is shorthand for $x + a \hat{e}_\mu$, where $\hat{e}_\mu$ is a Euclidean basis vector. From a geometrical viewpoint, gauge links define parallel transport along the edges of the lattice. Under general gauge transformations, gauge links transform according to
\begin{align}
    U'_{x,\mu} = \Omega_x U_{x,\mu} \Omega^\dagger_{x+\mu}.
\end{align}
Reversing the path yields the inverse link which we denote by $U^\dagger_{x,\mu} = U_{x+\mu, -\mu}$. Links can be concatenated to form plaquettes ($1 \times 1$ loops)
\begin{align}
    U_ {x,\mu\nu} = U_{x,\mu} U_{x+\mu, \nu} U_{x+\mu+\nu, -\mu} U_{x+\mu, -\mu}, \qquad 
    U'_{x,\mu\nu} = \Omega_x U_{x,\mu\nu} \Omega^\dagger_x,
\end{align}
which transform locally under lattice gauge transformations. The Wilson action is given by
\begin{align} \label{eq:wilson_action}
S_W[U] = \frac{2}{g^2} \sum_{x \in \Lambda} \sum_{\mu < \nu} \Tr \left[ \one - U_{x,\mu\nu} \right],
\end{align}
where $g > 0$ is the Yang-Mills coupling constant.

In these proceedings we review lattice gauge equivariant convolutional neural networks (L-CNNs), which is an equivariant neural network architecture proposed by the authors \cite{Favoni:2020reg} specifically tailored to $\SU(N_c)$ lattice gauge theory. This architecture allows for neural networks that use link variables in the input layer and satisfy equivariance under general lattice gauge transformations. We discuss some aspects of L-CNNs in detail and show the main results of our computational experiments, where we compare traditional (non-equivariant) CNNs to L-CNNs in specific regression tasks. 

\section{Lattice gauge equivariant convolutional neural networks}

The idea behind L-CNNs is to formulate CNNs which can be used to process gauge link configurations $\mathcal{U} = \{ U_{x,\mu} \}$, i.e.~the set of all gauge links on the lattice, in a gauge equivariant manner. In this section we review a few aspects of L-CNNs; a complete description can be found in our original paper \cite{Favoni:2020reg}. Adopting a similar notation as in \cite{Cohen:2016aaa}, a gauge equivariant function $g$ satisfies the equivariance property
\begin{align}
    T_\Omega g(\mathcal{U} ) = g(T_\Omega \mathcal{U} ),
\end{align}
where $T_\Omega$ denotes the application of a general gauge transformation $\Omega_x$. In the case of gauge links we have $T_\Omega U_{x,\mu} = \Omega_x U_{x,\mu} \Omega^\dagger_{x+\mu}$. The transformation properties of $T_\Omega g(\mathcal{U})$ generally do not need to be the same as the one for links. For example, the function $g(\mathcal{U})_x = U_{x,\mu\nu}$, which maps links to plaquettes, transforms locally
\begin{align}
    T_\Omega g(\mathcal{U})_x =  g(T_\Omega \mathcal{U})_x = \Omega_x U_{x,\mu\nu} \Omega^\dagger_x = \Omega_x g(\mathcal{U})_x \Omega^\dagger_x.
\end{align}
In particular, a function is gauge invariant if $T_\Omega g = g$, i.e.~if $T_\Omega$ acts as the identity map.

\begin{figure}
    \centering
    \begin{subfigure}[b]{0.25\textwidth}
    \centering
    \includegraphics[scale=1.2]{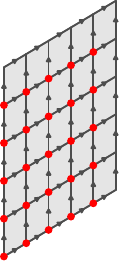}
    \caption{Lattice data $(\mathcal{U}, \mathcal{W})$}
    \label{fig:lcnn_input}
    \end{subfigure}
    \begin{subfigure}[b]{0.29\textwidth}
    \centering
    \includegraphics[scale=1.2]{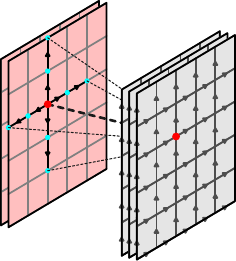}
    \caption{L-Conv layer}
    \label{fig:lconv}
    \end{subfigure}
    \begin{subfigure}[b]{0.37\textwidth}
    \centering
    \includegraphics[scale=1.2]{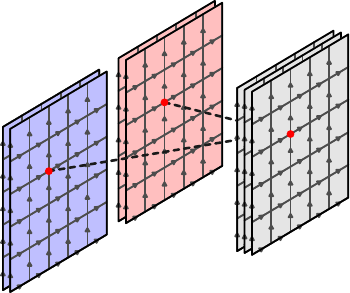}
    \caption{L-Bilin layer}
    \label{fig:lbilin}
    \end{subfigure}
    \caption{Schematic overview of different L-CNN layers. (a) The data $(\mathcal{U}, \mathcal{W})$ considered in L-CNNs is a tuple consisting of links $\mathcal{U}$ (gray edges) and locally transforming matrices $\mathcal{W}$ (red dots). (b) A lattice gauge equivariant convolution computes a weighted sum of parallel transported data (cyan points) along straight lines (black lines) up to a certain kernel size (here, $K=2$). (c) A lattice gauge equivariant bilinear layer multiplies data at a common lattice site (red dot), combining different channels in the process (blue and red planes). } 
    \label{fig:lcnn_fig1}
\end{figure}

More generally, we consider functions of the form $g(\mathcal{U}, \mathcal{W})$ acting on a tuple $(\mathcal{U}, \mathcal{W})$ consisting of the set of links and locally transforming matrices $\mathcal{W} = \{ W_{x,i} \}$ where $W_{x,i} \in \mathbb{C}^{N_c \times N_c}$ are general complex matrices and $i \in \{ 1, 2, \dots, N_\mathrm{ch} \}$ denotes different channels similar to traditional CNNs. We require that $\mathcal{W}$ variables transform locally: $T_\Omega W_{x,i} = \Omega_x W_{x,i} \Omega^\dagger_{x}$. For example, the set $\mathcal{W}$ may consist of all possible plaquettes on the lattice formed by links, but generally $\mathcal{W}$ variables can be considered independent of the links $\mathcal{U}$. A gauge equivariant function then satisfies
\begin{align}
    T_\Omega g(\mathcal{U}, \mathcal{W}) = g(T_\Omega \mathcal{U}, T_\Omega \mathcal{W}).
\end{align}
We formulate L-CNNs as CNNs consisting of individual layers which satisfy gauge equivariance. The idea behind the use of tuples $(\mathcal{U}, \mathcal{W})$ in L-CNNs is to explicitly split the input of the L-CNN into a feature map (or data) $\mathcal{W}$ and  link variables $\mathcal{U}$, which encode the geometrical information of how data at different lattice sites can be compared using parallel transport in a manner consistent with gauge equivariance. This approach is comparable to gauge equivariant CNNs \cite{Cohen:2019aaa} where the base manifold is the ``stage'' on which feature maps (which can be scalar-, vector- or tensor-valued) are defined, and the connection provides the necessary information to compare feature maps at different points.
Similarly, the gauge links $\mathcal{U}$ provide the stage for our feature maps $\mathcal{W}$. Figure \ref{fig:lcnn_input} shows the data in L-CNNs schematically. 

We exemplify the L-CNN using two particularly important layers: gauge equivariant convolutions  and bilinear layers.

\paragraph{Lattice gauge equivariant convolutions (L-Convs)}

Convolutional layers in CNNs combine data at different points by computing a sum of data of a feature map, where each term is weighted by a (trainable) kernel. Consider a two-dimensional, single-channel feature map $f: \mathbb{Z}^2 \rightarrow \mathbb{R}$, then a convolution can be defined as
\begin{align}
    f'_x = \sum_{y\in\mathbb{Z}^2} \omega_{x-y} f_y,
\end{align}
where $\omega_{x-y} \in \mathbb{R}$ are the kernel weights and $f'$ is the new feature map after the convolution.  The largest distance considered in a convolution defines the size $K \in \mathbb{N}$ of the kernel, i.e.~we assume the kernel to be compact. Computing a standard convolution of $\mathcal{W}$ variables would violate the requirement of gauge equivariance, unless one accounts for parallel transport:
\begin{align}
    W'_{x,i} = \sum_y \omega_{x-y} U_{x \leftarrow y} W_{y,i} U^\dagger_{x \leftarrow y},
\end{align}
where $U_{x \leftarrow y}$ is a Wilson line on the lattice tracing a path from $x$ to $y$. However, the choice of the path on the lattice is not unique. Whereas in the continuum one may only consider straight-line paths (or more generally geodesics), there are multiple shortest paths connecting $x$ and $y$ on the lattice. A possible solution is to only take straight paths along the lattice axes into account. Thus, we define the lattice gauge equivariant convolution (L-Conv) as 
\begin{align} \label{eq:lconv}
W'_{x, i} = \sum_{j, \mu, k} \omega_{i, j, \mu, k} U_{x, k\cdot \mu} W_{x + k\cdot \mu, j} U^\dagger_{x, k\cdot \mu},
\end{align}
where $\omega_{i, j, \mu, k} \in \mathbb{C}$ is a trainable kernel with $1 \le i \le N_\mathrm{ch,out}$, $1 \le j \le N_\mathrm{ch,in}$, $0 \le \mu \le D$ and \mbox{$-K$}$ \le k \le K$, where $K$ is the kernel size. The parallel transport from $x$ to $x+k\cdot \mu$ is given by
\begin{align}
    U_{x,k\cdot\mu} = \prod^{k-1}_{i=0} U_{x+i \cdot \mu, \mu} = U_{x,\mu} U_{x+\mu, \mu} U_{x+2\cdot\mu,\mu} \dots U_{x+(k-1)\cdot \mu, \mu}.
\end{align}
We allow for possible mixing of channels, and the number of channels can change from $N_\mathrm{ch,in}$ to $N_\mathrm{ch,out}$. Additionally, one may enlarge the channels of $\mathcal{W}$ by unit matrices prior to computing the convolution via
\begin{align}
     (W_{x,1}, W_{x,2}, \dots, W_{x,N_\mathrm{ch,in}}) \rightarrow (\one, W_{x,1}, W_{x,2}, \dots, W_{x,N_\mathrm{ch,in}}),
\end{align}
to allow for a bias term. Even more expressivity is gained by also including hermitian conjugates
\begin{align}
     (W_{x,1}, W_{x,2}, \dots, W_{x,N_\mathrm{ch,in}}) \rightarrow (\one, W_{x,1}, W_{x,2}, \dots, W_{x,N_\mathrm{ch,in}}, W^\dagger_{x,1}, W^\dagger_{x,2}, \dots, W^\dagger_{x,N_\mathrm{ch,in}}).
\end{align}
We note that L-Convs, similar to any convolutional layer, are equivariant under translations of the lattice. Figure \ref{fig:lconv} visualizes the L-Conv layer.

\paragraph{Lattice gauge equivariant bilinear layers (L-Bilin)}

Parallel transport is required when comparing data at different lattice sites, but one can easily define operations acting only on single points without violating gauge equivariance.
%Furthermore, neural networks typically require non-linear layers, such as activation functions, to be able to express arbitrary non-linear functions.
An example for such a layer in the L-CNN  is the lattice gauge equivariant bilinear layer (L-Bilin) which combines two tuples $(\mathcal{U}, \mathcal{W})$ and $(\mathcal{U}, \mathcal{W'})$ into a single tuple $(\mathcal{U}, \mathcal{W}'')$ in a bilinear manner. We define L-Bilin operations via
\begin{align} \label{eq:lbilin}
W''_{x,i} = \sum_{j,k} \alpha_{i,j,k} W_{x,j} W'_{x,k},
\end{align}
where $\alpha_{i,j,k} \in \mathbb C$ are trainable weights with $1 \le i \le N_\mathrm{out}$, $1 \le j \le N_\mathrm{in, 1}$ and $1 \le k \le N_\mathrm{in, 2}$. The product of the two matrices $W_{x,j}$ and $W'_{x,k}$ yields a matrix that transforms locally at $x$ and is thus compatible with gauge equivariance. L-Bilin is linear in each of the two arguments $\mathcal{W}$ and $\mathcal{W}'$, however one may also choose $\mathcal{W}' = \mathcal{W}$ yielding a quadratic operation. Similar to L-Convs, the channels of $\mathcal{W}$ can be extended by unit matrices and hermitian conjugates. This way the bilinear layer contains a bias term and linear terms as well (through multiplication with unit elements of the other argument).

\paragraph{Other layers and the expressivity of L-CNNs}

\begin{figure}
    \centering
    \includegraphics{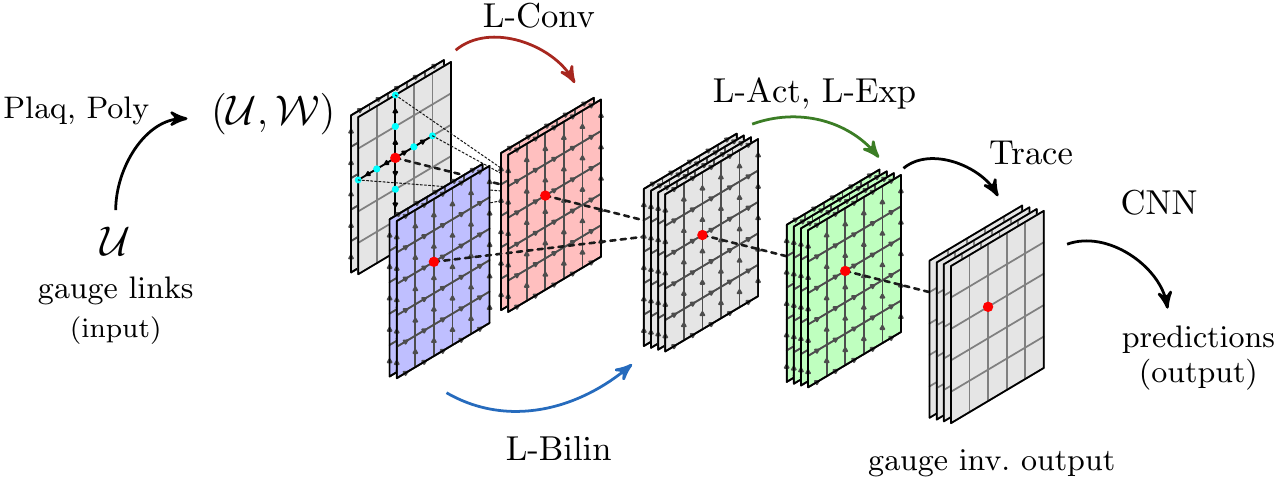}
    \caption{Schematic of a feed-forward L-CNN consisting of multiple layers. Links are first pre-processed to generate plaquettes (and/or Polyakov loops) before being fed to L-Conv and L-Bilin layers. Activation functions and exponentiation layers may also be used. For gauge invariant predictions, a Trace layer is required. Figure adapted from \cite{Favoni:2020reg}.}
    \label{fig:lcnn}
\end{figure}

Our paper \cite{Favoni:2020reg} proposes more types of layers, which we only briefly summarize here. For example, one may combine an L-Conv and an L-Bilin layer into a single operation, which we found to be easier to train than separate convolutions and bilinear layers.
Other types are lattice gauge equivariant activation functions (L-Act), which locally multiply $\mathcal{W}$ variables with gauge invariant scalars, and lattice gauge equivariant exponentiation layers (L-Exp), which can modify the set of gauge links $\mathcal{U}$. Another important layer is the Trace layer, which maps $(\mathcal{U}, \mathcal{W})$ to gauge invariant variables $\mathcal{T}_{x,i}$ via
\begin{align}
\mathcal{T}_{x,i} = \mathrm{Tr} \left[ W_{x,i} \right].
\end{align}
%Note that this layer does not contain any trainable weights. Due to the trace of locally transforming matrices, the output of this layer is invariant under gauge transformations.
This layer is used when the output of the L-CNN is required to be invariant, e.g.~if the L-CNN is used to fit invariant observables.  To keep the discussion as general as possible, we have not yet specified what information the $\mathcal{W}$ variables should contain, except requiring them to transform locally. As a pre-processing step to any L-CNN, one may generate $\mathcal{W}$ from the set of gauge links $\mathcal{U}$ by computing all plaquettes (which we refer to as a Plaq layer) and Polyakov loops (Poly layer). 

Generic L-CNNs are comprised of multiple layers, as shown in Fig.~\ref{fig:lcnn}, similar to standard CNNs. It is well known that deep CNNs can describe arbitrary continuous functions (see e.g.~universality theorems in \cite{Zhou:2018a}). The question then arises as to what class of functions can be represented using L-CNNs. In \cite{Favoni:2020reg} we show that by stacking multiple L-Conv and L-Bilin layers and using plaquettes as input $\mathcal{W}$ variables, L-CNNs can generate contractible Wilson loops of arbitrary shape and size. Non-contractible Wilson loops can also be generated by including Polyakov loops in the input layer. Similar to universality theorems for deep CNNs, the use of L-Act layers allows arbitrary non-linear gauge equivariant functions to be expressed as L-CNNs. 

\section{Computational experiments} \label{sec:experiments}

\begin{figure}
    \centering
    \includegraphics[width=\textwidth]{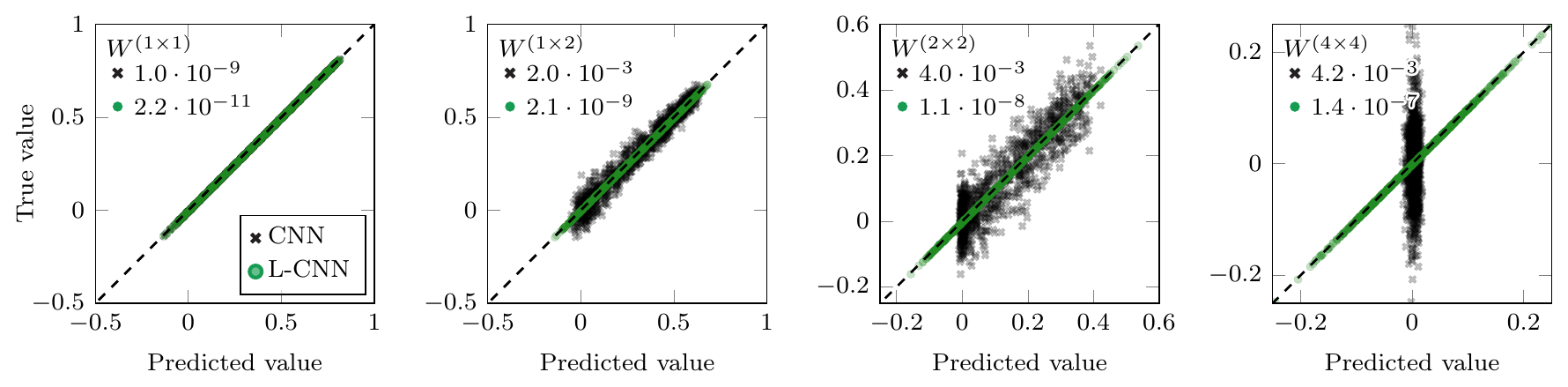}
    \caption{Scatter plot of our best L-CNNs (green circles) vs.~our best CNNs (black crosses) on $8 \times 8$ test data for various sizes of the Wilson loop. The dashed line indicates perfect agreement between prediction and true value. MSEs are stated in the top left corner of each panel. Figure from \cite{Favoni:2020reg}.}
    \label{fig:scatter}
\end{figure}

In order to test the L-CNN framework in practice, we study a series of seemingly simple regression problems. Restricting the base space to a quadratic two dimensional lattice ($8 \times 8$ up to $64 \times 64$) and focusing on $\SU(2)$, we generate gauge field configurations from a Markov Chain  Monte Carlo simulation at various values of the coupling constant. These configurations are split into separate training, validation and test datasets. For each individual configuration, we compute the real value of traced Wilson loops of various sizes:
\begin{align}
    W^{(m \times n)}_{x,\mu\nu} = \frac{1}{N_c} \mathrm{Re}\,  \Tr \left[ U^{(m \times n)}_{x,\mu\nu} \right],
\end{align}
where $n, m \geq 1$ define the width and height of the loop. We generate loops of size $1 \times 1$ (plaquettes), $1 \times 2$, $2 \times 2$ and $4 \times 4$. We then formulate a regression problem where the lattice configurations should be mapped to the value of a particular traced loop using different L-CNN and non-equivariant CNN architectures, i.e.~the input is given by the links $\mathcal{U}$ and the desired output (labels) are $W^{(m \times n)}_{x,\mu\nu}$. L-CNN architectures consist of multiple L-Conv+L-Bilin layers and a Trace layer at the end, similar to Fig.~\ref{fig:lcnn}. For comparison, we study a wide range of different non-equivariant CNN architectures of various widths, depths and employing different activation functions. The CNNs are provided with the same input as the L-CNN after pre-processing, namely links and plaquettes. Both types of architectures are implemented and trained using the \textit{PyTorch} framework. Networks are optimized by minimizing the mean squared error (MSE) with respect to trainable weight parameters. Training and validation are performed on the smallest lattice size ($8 \times 8$), but models are tested also on larger lattice sizes.

Figure \ref{fig:scatter} shows one of the main results of this study, where the predictions of an L-CNN and a CNN are plotted against the true values (labels) of the test dataset. As we have trained multiple networks of both types (100 L-CNNs, 2680 CNNs), we only show the best models  encountered in our study, which are selected using MSE on the validation dataset. We observe that L-CNN architectures are able to make sensible predictions in all cases. On the other hand, the performance of non-equivariant CNNs starts to deteriorate with larger loop size. In the case of the largest loops, CNNs seem to find no correlation between input and output at all, which is indicated by an almost constant prediction in the scatter plot (rightmost panel in Fig.~\ref{fig:scatter}). Furthermore, even our best CNNs are sensitive to gauge transformations, i.e.~the output of a non-equivariant CNN can change drastically when processing a gauge equivalent lattice configuration, whereas L-CNNs are gauge invariant by construction. We have also demonstrated that L-CNNs are able to solve similar regression problems on four-dimensional lattices (up to $4 \times 4$ Wilson loops on $8\cdot 16^3$ lattices). We refer the reader to the original paper \cite{Favoni:2020reg} for more details. Our code and datasets for these computational experiments can be found in our GitLab repository\footnote{See \href{https://gitlab.com/openpixi/lge-cnn/}{https://gitlab.com/openpixi/lge-cnn/}.
% https://gitlab.com/openpixi/lge-cnn/
}.

\section{Summary and outlook}

In these proceedings we have highlighted some aspects of the L-CNN architecture with a particular focus on the concept of gauge equivariance and motivating the formulation of L-Conv and L-Bilin layers. The L-CNN is a general, genuinely gauge equivariant neural network that can be used for generic machine learning tasks in $\SU(N_c)$ lattice gauge theory. In our computational experiments we have demonstrated that L-CNNs are able to solve non-linear regression problems, where non-equivariant CNNs fail to make sensible predictions.

Up until now we have not made use of (or implemented) L-Act and L-Exp layers, as our main goal was to establish that L-CNNs can be used to generate arbitrary loops, which requires only L-Conv and L-Bilin layers. Although we have only tested $\SU(2)$, our implementation works for any $\SU(N_c)$ gauge group. Using slight modifications, the code could also be adapted for U$(1)$. However, the L-CNN code requires large computational resources during training, in particular with regards to memory consumption, and a more efficient implementation would be desirable. This is of particular importance when using larger lattice sizes in four dimensions or studying more complicated problems.

L-CNNs have been explicitly formulated for $\SU(N_c)$ lattice gauge theory, which is a discretization of (continuum) non-Abelian gauge theory. Although such a lattice formulation is required to perform practical computations, a more general mathematical formulation in terms of principal bundles (similar to gauge equivariant CNNs \cite{Cohen:2019aaa}) would be desirable from a theoretical viewpoint. Putting the L-CNN on better theoretical foundations could help us generalize L-CNNs to other types of data such as fermionic fields and develop new equivariant layers.

Our computational experiments focused on toy model regression problems, which we conducted to compare the performance of L-CNNs and CNNs in a transparent way. It would be interesting to see how L-CNNs (in particular L-Convs) could be applied to more practical problems in e.g.~normalizing flow models \cite{Kanwar:2020xzo, Boyda:2020hsi, Albergo:2021vyo} or models that modify gauge links \cite{Tomiya:2021ywc}. Another exciting direction would be to use L-CNNs to formulate gauge equivariant continuous flows \cite{deHaan:2021erb} to potentially speed up the generation of lattice configurations for independent Monte Carlo sampling, while retaining a large degree of symmetry. A related problem would be to use L-CNNs to upscale real-time simulations of classical lattice gauge fields used in the early stages of relativistic heavy-ion collisions \cite{Ipp:2020igo}.

\acknowledgments
This work has been supported by the Austrian Science Fund FWF No.~P32446-N27, No.~P28352 and Doctoral program No. W1252-N27. The Titan\,V GPU used for this research was donated by the NVIDIA Corporation.

\bibliographystyle{JHEP.bst}
\bibliography{refs.bib}

\providecommand{\href}[2]{#2}\begingroup\raggedright\begin{thebibliography}{10}

\bibitem{Noether:1918}
E.~Noether, \emph{{Invariante Variationsprobleme}}, {\emph{Nachrichten von der
  Gesellschaft der Wissenschaften zu G\"ottingen, Mathematisch-Physikalische
  Klasse} {\bfseries 1918} (1918) 235}.

\bibitem{Bronstein:2021aaa}
M.M.~{Bronstein}, J.~{Bruna}, T.~{Cohen} and P.~{Veli{\v{c}}kovi{\'c}},
  \emph{{Geometric Deep Learning: Grids, Groups, Graphs, Geodesics, and
  Gauges}},  \href{https://arxiv.org/abs/2104.13478}{{\ttfamily 2104.13478}}.

\bibitem{Gerken:2021sla}
J.E.~Gerken, J.~Aronsson, O.~Carlsson, H.~Linander, F.~Ohlsson, C.~Petersson
  et~al., \emph{{Geometric Deep Learning and Equivariant Neural Networks}},
  \href{https://arxiv.org/abs/2105.13926}{{\ttfamily 2105.13926}}.

\bibitem{Bulusu:2021rqz}
S.~Bulusu, M.~Favoni, A.~Ipp, D.I.~M\"uller and D.~Schuh, \emph{{Generalization
  capabilities of translationally equivariant neural networks}},
  \href{https://doi.org/10.1103/PhysRevD.104.074504}{\emph{Phys. Rev. D}
  {\bfseries 104} (2021) 074504}
  [\href{https://arxiv.org/abs/2103.14686}{{\ttfamily 2103.14686}}].

\bibitem{Cohen:2016aaa}
T.S.~Cohen and M.~Welling, \emph{Group equivariant convolutional networks},  in
  \emph{Proceedings of The 33rd International Conference on Machine Learning},
  vol.~48, pp.~2990--2999, JMLR, Jun, 2016
  [\href{https://arxiv.org/abs/1602.07576}{{\ttfamily 1602.07576}}].

\bibitem{Cohen:2019aaa}
T.S.~Cohen, M.~Weiler, B.~Kicanaoglu and M.~Welling, \emph{Gauge equivariant
  convolutional networks and the icosahedral {CNN}},  in \emph{Proceedings of
  the 36th International Conference on Machine Learning}, vol.~97,
  pp.~1321--1330, JMLR, Jun, 2019
  [\href{https://arxiv.org/abs/1902.04615}{{\ttfamily 1902.04615}}].

\bibitem{Kanwar:2020xzo}
G.~Kanwar, M.S.~Albergo, D.~Boyda, K.~Cranmer, D.C.~Hackett, S.~Racani\`ere
  et~al., \emph{{Equivariant flow-based sampling for lattice gauge theory}},
  \href{https://doi.org/10.1103/PhysRevLett.125.121601}{\emph{Phys. Rev. Lett.}
  {\bfseries 125} (2020) 121601}
  [\href{https://arxiv.org/abs/2003.06413}{{\ttfamily 2003.06413}}].

\bibitem{Boyda:2020hsi}
D.~Boyda, G.~Kanwar, S.~Racani\`ere, D.J.~Rezende, M.S.~Albergo, K.~Cranmer
  et~al., \emph{{Sampling using $SU(N)$ gauge equivariant flows}},
  \href{https://doi.org/10.1103/PhysRevD.103.074504}{\emph{Phys. Rev. D}
  {\bfseries 103} (2021) 074504}
  [\href{https://arxiv.org/abs/2008.05456}{{\ttfamily 2008.05456}}].

\bibitem{Albergo:2021vyo}
M.S.~Albergo, D.~Boyda, D.C.~Hackett, G.~Kanwar, K.~Cranmer, S.~Racani\`ere
  et~al., \emph{{Introduction to Normalizing Flows for Lattice Field Theory}},
  \href{https://arxiv.org/abs/2101.08176}{{\ttfamily 2101.08176}}.

\bibitem{Tomiya:2021ywc}
A.~Tomiya and Y.~Nagai, \emph{{Gauge covariant neural network for 4 dimensional
  non-abelian gauge theory}},
  \href{https://arxiv.org/abs/2103.11965}{{\ttfamily 2103.11965}}.

\bibitem{Wilson:1974sk}
K.G.~Wilson, \emph{Confinement of quarks},
  \href{https://doi.org/10.1103/PhysRevD.10.2445}{\emph{Phys. Rev. D}
  {\bfseries 10} (1974) 2445}.

\bibitem{Favoni:2020reg}
M.~Favoni, A.~Ipp, D.I.~M\"uller and D.~Schuh, \emph{{Lattice gauge equivariant
  convolutional neural networks}},
  \href{https://arxiv.org/abs/2012.12901}{{\ttfamily 2012.12901}}.

\bibitem{Zhou:2018a}
D.-X.~Zhou, \emph{Universality of deep convolutional neural networks},
  \href{https://doi.org/https://doi.org/10.1016/j.acha.2019.06.004}{\emph{Applied
  and Computational Harmonic Analysis} {\bfseries 48} (2020) 787}
  [\href{https://arxiv.org/abs/1805.10769}{{\ttfamily 1805.10769}}].

\bibitem{deHaan:2021erb}
P.~de~Haan, C.~Rainone, M.~Cheng and R.~Bondesan, \emph{{Scaling Up Machine
  Learning For Quantum Field Theory with Equivariant Continuous Flows}},
  \href{https://arxiv.org/abs/2110.02673}{{\ttfamily 2110.02673}}.

\bibitem{Ipp:2020igo}
A.~Ipp and D.I.~M\"uller, \emph{{Progress on 3+1D Glasma simulations}},
  \href{https://doi.org/10.1140/epja/s10050-020-00241-6}{\emph{Eur. Phys. J. A}
  {\bfseries 56} (2020) 243}
  [\href{https://arxiv.org/abs/2009.02044}{{\ttfamily 2009.02044}}].

\end{thebibliography}\endgroup

\end{document}